\newcommand{\be}{\begin{equation}}
\newcommand{\ee}{\end{equation}}
\newcommand{\ba}{\begin{eqnarray}}
\newcommand{\ea}{\end{eqnarray}}
\newcommand{\bi}{\begin{itemize}}
\newcommand{\ei}{\end{itemize}}

\newcommand{\<}{\langle} 
\renewcommand{\>}{\rangle}  
\newcommand{\eq}{Eq.~}

\newcommand{\fig}{Fig.~}

\newcommand{\la}{\label}
\newcommand{\rmi}[1]{{\mbox{\scriptsize #1}}}
\newcommand{\rmO}{{\mathcal{O}}}

\newcommand{\rmii}[1]{{\mbox{\tiny\rm{#1}}}}

\newcommand{\im}{\mathop{\mbox{Im}}}
\newcommand{\Tint}[1]{{\hbox{$\sum$}\!\!\!\!\!\!\!\int\,}_{\!\!\!\!\raise-0.9ex\hbox{$\scriptstyle{#1}$}}}
\newcommand{\Tinti}[1]{{{\Sigma}\!\!\!\!\raise0.3ex\hbox{$\int$}_\rmii{${#1}$}}}

\documentclass[3p,times]{elsarticle}

\usepackage{ecrc}

\usepackage{epstopdf}

\volume{00}

\firstpage{1}

\journalname{Nuclear Physics A}

\runauth{H.B.\ Meyer et al.}

\jid{nupha}

\jnltitlelogo{Nuclear Physics A}

\usepackage{graphicx}
\usepackage{amsmath,amssymb}

\begin{document}

\begin{frontmatter}



\title{Vector screening masses in the quark-gluon plasma\\ and their physical significance}



\author[Regensburg]{B.\ B.\ Brandt}
\author[Mainz]{A.\ Francis}
\author[Bern]{M.\ Laine}
\author[Mainz]{H.\ B.\ Meyer}

\address[Regensburg]{Institute for Theoretical Physics, 
        University of Regensburg,  
        93040 Regensburg, Germany}
\address[Mainz]{PRISMA Cluster of Excellence,
Institut f\"ur Kernphysik and Helmholtz~Institut~Mainz,
J.\ Gutenberg-Universit\"at Mainz, D-55099 Mainz, Germany}
\address[Bern]{Institute for Theoretical Physics, Albert Einstein Center, University of Bern, 
        Sidlerstrasse 5, 3012 Bern, Switzerland}

\begin{abstract}
Static and non-static thermal screening states that couple to the
conserved vector current are investigated in the high-temperature
phase of QCD. Their masses and couplings to the current are determined
at weak coupling, as well as using two-flavor lattice QCD
simulations. A consistent picture emerges from the comparison,
providing evidence that non-static Matsubara modes can indeed be
treated perturbatively. We elaborate on the physical significance of
the screening masses.
\end{abstract}

\begin{keyword}
Screening Masses \sep Lattice QCD \sep Resummed Perturbation Theory

\end{keyword}

\end{frontmatter}



\section{Introduction}
\label{intro}

While lattice QCD is a primary tool to investigate equilibrium
properties of finite-temperature QCD, real-time quantities, such as
transport coefficients or production rates, are notoriously difficult
to address. In this work, we are motivated by looking for an indirect
way of probing real-time physics in the Euclidean formulation of
thermal field theory, not involving a numerically ill-posed analytic
continuation~\cite{Meyer:2011gj}. Specifically, we
show~\cite{Brandt:2014uda} that there is a class of Euclidean
observables, namely flavor non-singlet (mesonic) screening masses at
non-zero Matsubara frequency, which are sensitive to the same infrared
physics as is relevant for jet quenching or photon and dilepton
production.  Concretely, the link between non-static correlation
lengths and the photon production rate is established via a certain
potential $V^+$ introduced previously and recently computed
non-perturbatively using lattice simulations
\cite{CaronHuot:2008ni,Panero:2013pla}.

\section{Calculation of vector screening masses}

\begin{figure}
\begin{center}
\includegraphics[width=5.9cm]{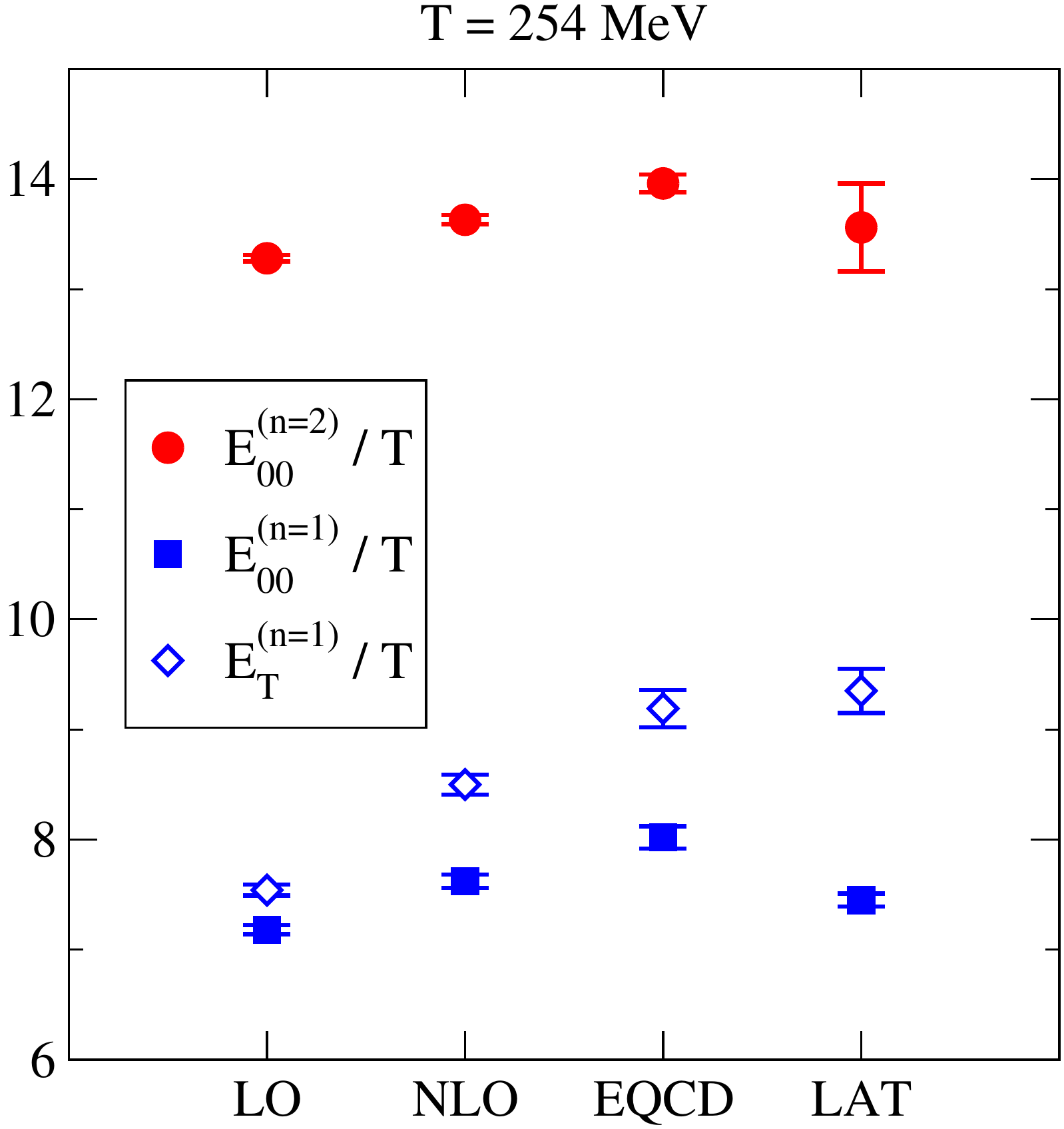}~~~~~~~
\includegraphics[width=6.1cm]{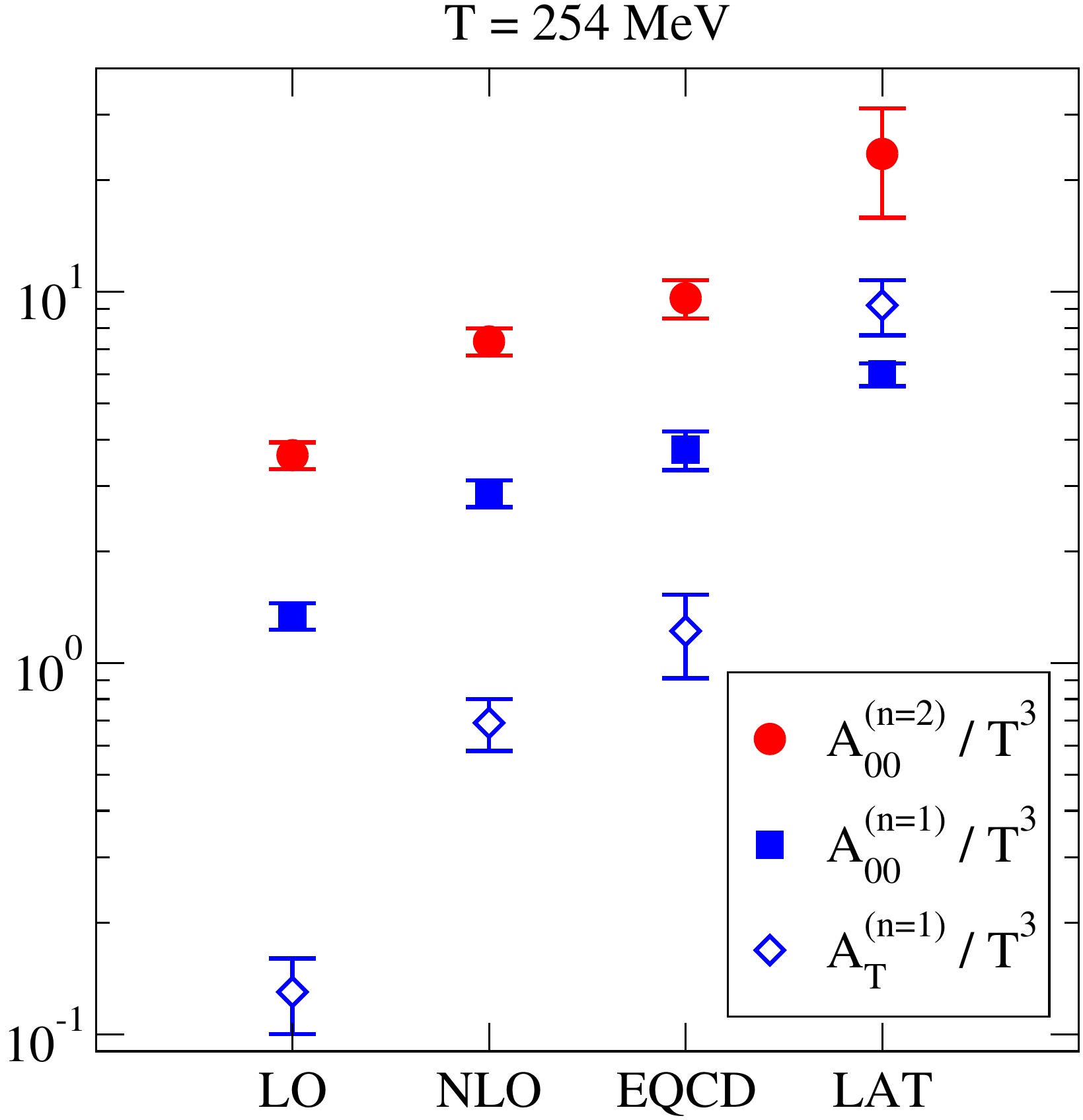}
\caption{Leading-order, next-to-leading order and non-perturbative EQCD potential: comparison of the predictions
to lattice data for the energy levels (left) and, on the right, the corresponding `amplitudes' (i.e.\ $-\frac{1}{\pi}\times$ 
the coefficient of the delta function in \eq(\ref{eq:rhokn})).}
\vspace{-0.5cm}
\label{fig:compa}
\end{center}
\end{figure}

We consider the  flavor non-singlet vector current correlator ($\vec{x}\equiv (x_1,x_2)$)
\be\la{eq:Gz}
 G^{(k_n)}_{\mu\nu}(z) \equiv
 \int_0^{1/T} \! {\rm d}\tau \, e^{i k_n \tau} \int_{\vec{x}}
 \Bigl\langle 
  V_\mu(\tau,\vec{x},z)  ~ V_\nu(0) 
 \Bigr\rangle_\rmi{c}
  \;  \stackrel{\mu = \nu}{=} \; 
  \int_0^\infty \! \frac{{\rm d}\omega}{\pi} \, 
  e^{-\omega |z|} \, \rho^{(k_n)}_{\mu\nu}(\omega)
  \;,\qquad \quad k_n \equiv 2\pi n T.
\ee
In the last equality, a spectral representation is given in terms of screening states.
We now build an effective theory (EFT) which allows us to describe
the physics of the correlators considered around the threshold
$\omega \sim \mathop{\mbox{max}}(k_1,k_n)$.
Dimensional reduction involves
keeping only the Matsubara zero modes of the SU(3) gauge fields 
in the covariant  derivatives $D_\mu = \partial_\mu - i g A_\mu$.
With $\psi^\dagger = \frac{1}{\sqrt{T}}\left( \chi^\dagger ~\;  \phi^\dagger\right)$, 
this leads to (at tree-level, in a certain representation of the $\gamma_\mu$'s)
\ba
 S^{ }_0 & = & 
 \sum_{ \{ p_n \} } \int_{\vec{x},z} 
 \biggl[
   i \chi^{\dagger}_{p_n} \biggl(p_n - g A_0 + D_3
   - \frac{D_iD_i + i \sigma_3 \epsilon_{ij} D_i D_j }{2p_n}
  \biggr) \,\chi_{p_n} 
 \\ & & \qquad 
 + \, 
   i \phi^{\dagger}_{p_n} \biggl(p_n - g A_0 - D_3 
   - \frac{D_iD_i + i \sigma_3 \epsilon_{ij} D_i D_j }{2p_n}
  \biggr) \,\phi_{p_n} 
 + \rmO\biggl( \frac{1}{p_n^2} \biggr)
  \biggr]
 \;,\qquad p_n \equiv 2\pi T \left( n + \frac{1}{2} \right) \;.
\nonumber
\ea
The fermions now have non-relativistic propagators and a simple inspection 
reveals that the `forward-propagating' mesons are represented by the fields 
$\phi^\dagger_{p_n} \chi_{p_n'}$
and $\phi^\dagger_{p_n} \phi_{-p_n'}$ with $p_n, p_n' > 0$. 
\begin{figure}
\begin{center}
\includegraphics[width=7.cm]{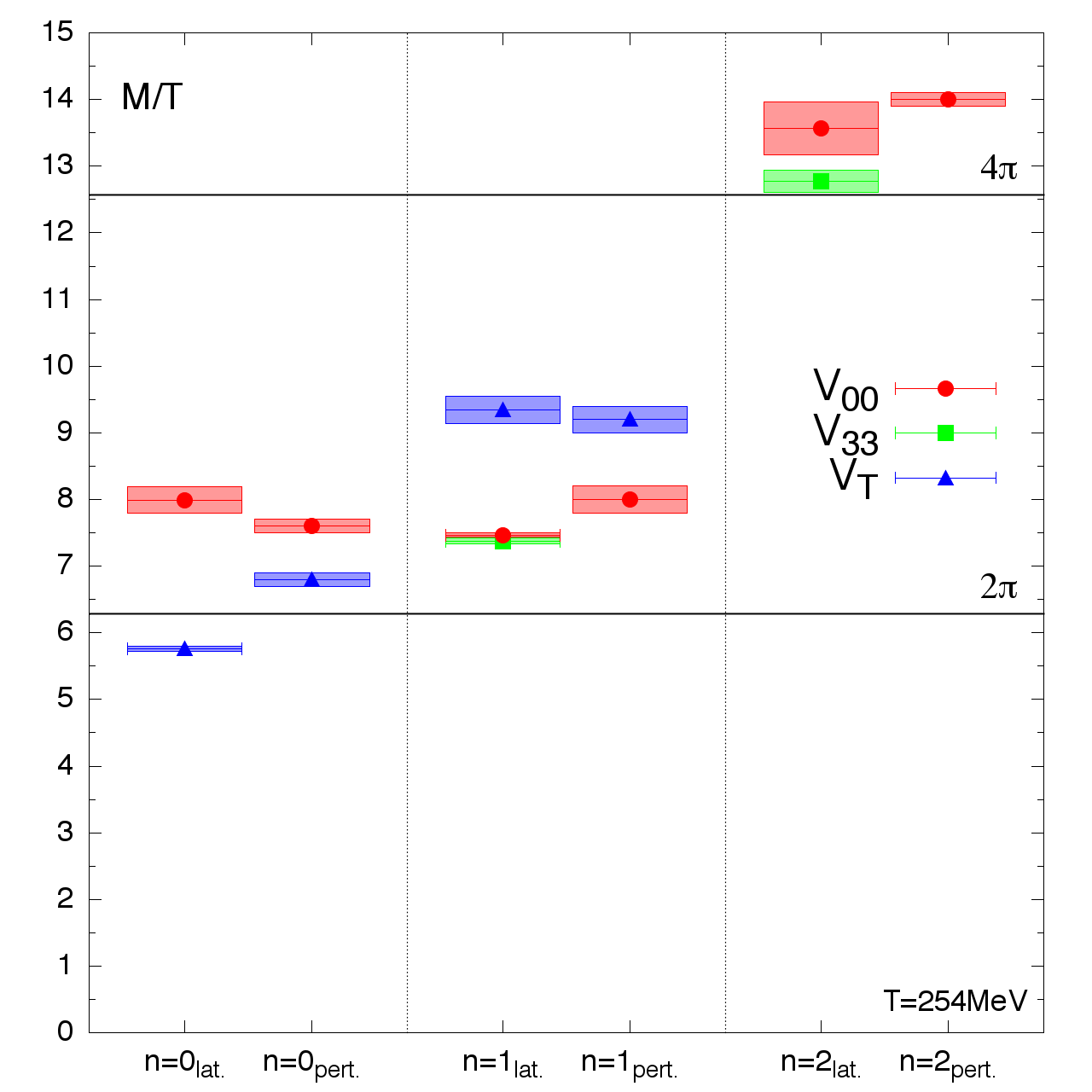}
\includegraphics[width=7.cm]{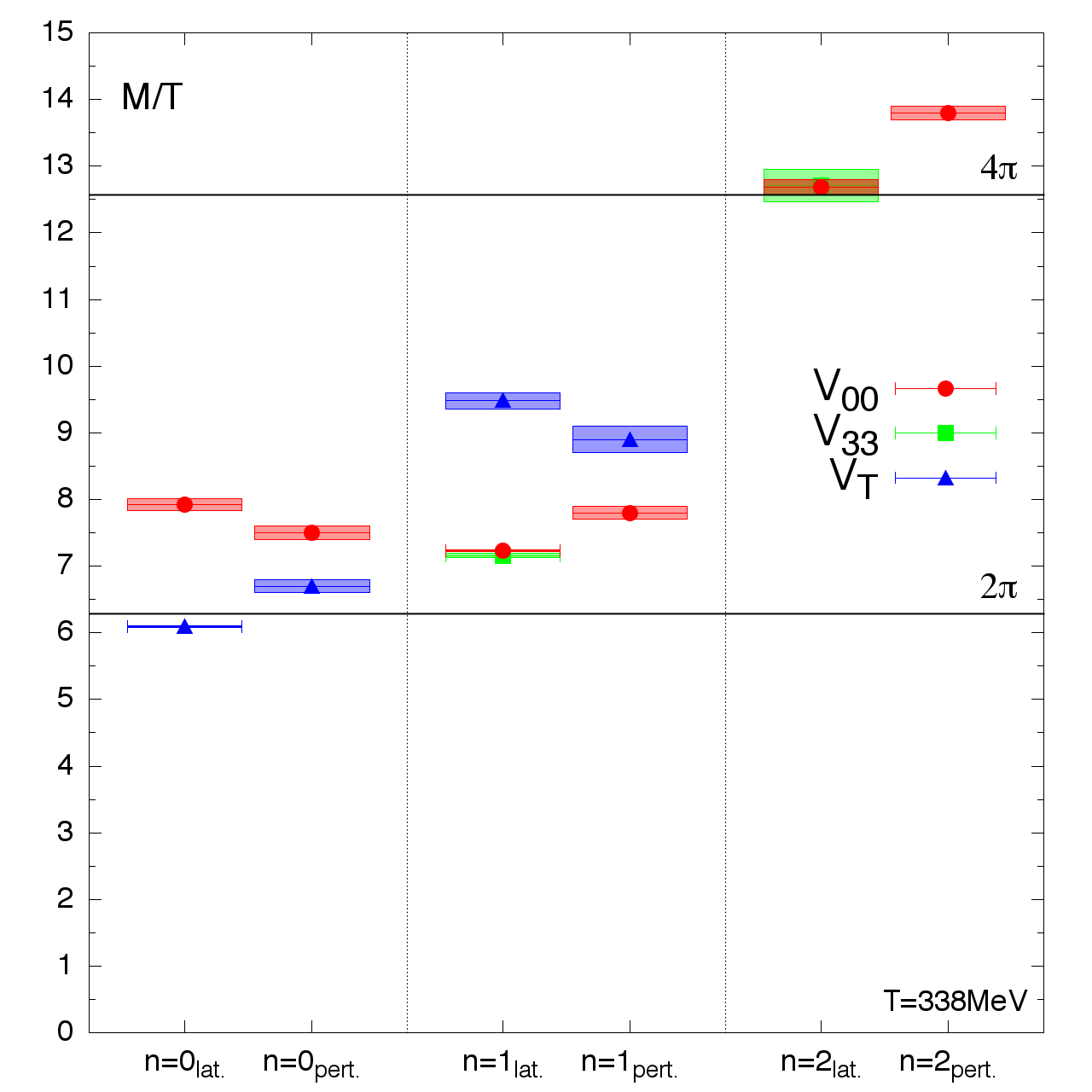}
\caption{Comparison of vector screening masses in two-flavor O($a$) improved Wilson lattice QCD 
simulations and in the EFT at two temperatures (corresponding to $16\times 64^3$ and $12\times 64^3$ lattices). 
The longitudinal and transverse channels are given for the Matsubara sectors $n=0,1$ and 2.}
\vspace{-0.3cm}
\label{fig:latvsEFT}
\end{center}
\end{figure}
For instance, consider for $k_n>0$
\be
 V_0^{(k_n)} = 
 \sum_{0 < p_n < k_n} \Big( 
   \chi^{\dagger}_{p_n } \chi_{p_n - k_n} + 
   \phi^{\dagger}_{p_n}  \phi_{p_n - k_n} 
  \Big)
 \;. 
\ee
For given $p_n$, let $w(z,\vec{y}\,)$ be the screening correlator of a local current at the source 
and a point-split current with separation $\vec{y}$ at the sink in the EFT. For $z>0$,
\ba
 (\partial_z + \hat{H}^{+})w(z,\vec{y}\,) = 0 \;, &\qquad& w(0,\vec y\,) = \delta^{(2)}(\vec{y}\,), 
\\
 \hat{H}^{+} \equiv  M_\rmi{cm} - \frac{\nabla^2}{2 M_\rmi{r}} + V^+ ,
&\qquad \quad &
 V^{+}_\rmi{LO}(\vec{y}\,) 
  =  \frac{ g_{E}^2 C_F }{2\pi} 
  \bigg[ 
    \ln\Big( \frac{m_E y }{2} \Big) + \gamma_{_{\rm E}} + K_0^{ }(m_E y )
  \bigg].
\ea
For $M_\rmi{cm}$, we perform a one-loop matching of the fermion masses 
in the reduced theory~\cite{Brandt:2014uda}, while $g_E^2=g^2T+\dots$ and 
$m_E^2=(\frac{N_c}{3}+\frac{N_f}{6}) g^2T^2$ is the Debye mass. The Fourier transform  of $w(z,\vec{y}\,)$
is closely related to the `resolvent' $g^+$, which obeys
$ ( \hat{H}^{+} -\omega - i 0^+ ) {g}^{+}(\omega,\vec{y}\, ) = \delta^{(2)}(\vec{y}\,) $
and is given explicitly by (the $\psi_i$ are normalized energy eigenstates)
\be
g^{+}(\omega,\vec{y}\,) = \sum_{i=0}^{\infty} 
 \frac{\psi^{ }_i(\vec{y}\,)\;\psi_i^*(\vec{0}\,)}{E_i - \omega - i 0^+}.
\ee
Finally, the screening spectral function can be obtained from the imaginary part of the resolvent,
\be\la{eq:rhokn}
 \rho^{(k_n)}_{00} (\omega) = - \sum_{0 < p_n < k_n} 2 N_c T
 \lim_{\vec{y}\to\vec{0}} \im {g}^{+}(\omega,\vec{y}\,) 
 = - 2 \pi N_c T\!\!\! \sum_{0 < p_n < k_n}
 \sum_{i=0}^{\infty} \delta(E_i - \omega)\; |\psi^{ }_i(\vec{0}\,)|^2
 \;.
\ee
We observe a close resemblance with the corresponding equations
appearing in the LPM resummation of longitudinal modes for photon or
dilepton production \cite{Aurenche:2002wq}: the same potential $V^+$
appears, and we are looking for an $s$-wave solution.  The
potential $V^+$ can be defined non-perturbatively using a (modified)
Wilson loop \cite{CaronHuot:2008ni} and has been computed in 3d
lattice simulations \cite{Panero:2013pla}.  The predictions for
non-static screening masses resulting from solving the Schr\"odinger
equation with the potential $V^+$ can be tested against direct
four-dimensional lattice QCD simulations; see \fig
\ref{fig:latvsEFT}, where also results for the transverse channel 
($\mu=\nu=1$ or 2 in \eq(\ref{eq:Gz})) are displayed.  
We consider the agreement to be satisfactory,
since the temperature is not very high. \fig \ref{fig:compa}
shows that using a non-perturbative potential, defined from a modified
Wilson loop, improves the overall agreement.

\section{Analytic continuation of the screening masses $E(\omega_n)$ in $\omega_n$}

\noindent Linear response along with a constitutive equation for the current $\vec V$
 makes a prediction for the retarded correlator of $V_0$ at
small frequency $\omega$ and wavevector $k$, 
\be\la{eq:linresp}
G_R(\omega,k) \stackrel{\omega,k\to0}{=\!=\!=} 
\frac{\chi_s D k^2}{-i\omega + D k^2},\qquad \quad \chi_s \equiv \int \!d^4\!x\; \< V_0(x)V_0(0)\>.
\ee
In terms of the frequency $\omega$, this corresponds to a pole at 
$\omega = -iD  k^2$ in the lower half of the complex plane, with $D$ the diffusion constant.
Continuing to imaginary frequencies, we obtain the Euclidean correlator,
$G_E(\omega_n,k) = G_R(i\omega_n,k) $ ($ k\neq 0;~ \omega_n = 2\pi T n >  0$).
One can also contemplate expression (\ref{eq:linresp}) from the point of view of $k$. 
Now, the poles in $k$ of $G_E(\omega_n,k)$ are equal to $i$ times a screening `mass' 
corresponding to a fixed Matsubara frequency sector $\omega_n$.
If we continue the screening mass as a function of the Matsubara frequency, 
Eq.\  (\ref{eq:linresp}) suggests that for $\omega_n\to 0$, we should find a $k$-pole at 
$k^2 = - \frac{\omega_n}{D}$, i.e.\  that one screening mass $E(\omega_n)$ continues to 
\be\la{eq:Esqrt}
E(\omega_n) \stackrel{\omega_n\to0}{\sim} \sqrt{{\omega_n}/{D}}.
\ee
This equation suggests a way to estimate the diffusion constant.

\begin{figure}
\begin{center}
\centerline{\includegraphics[width=8.cm]{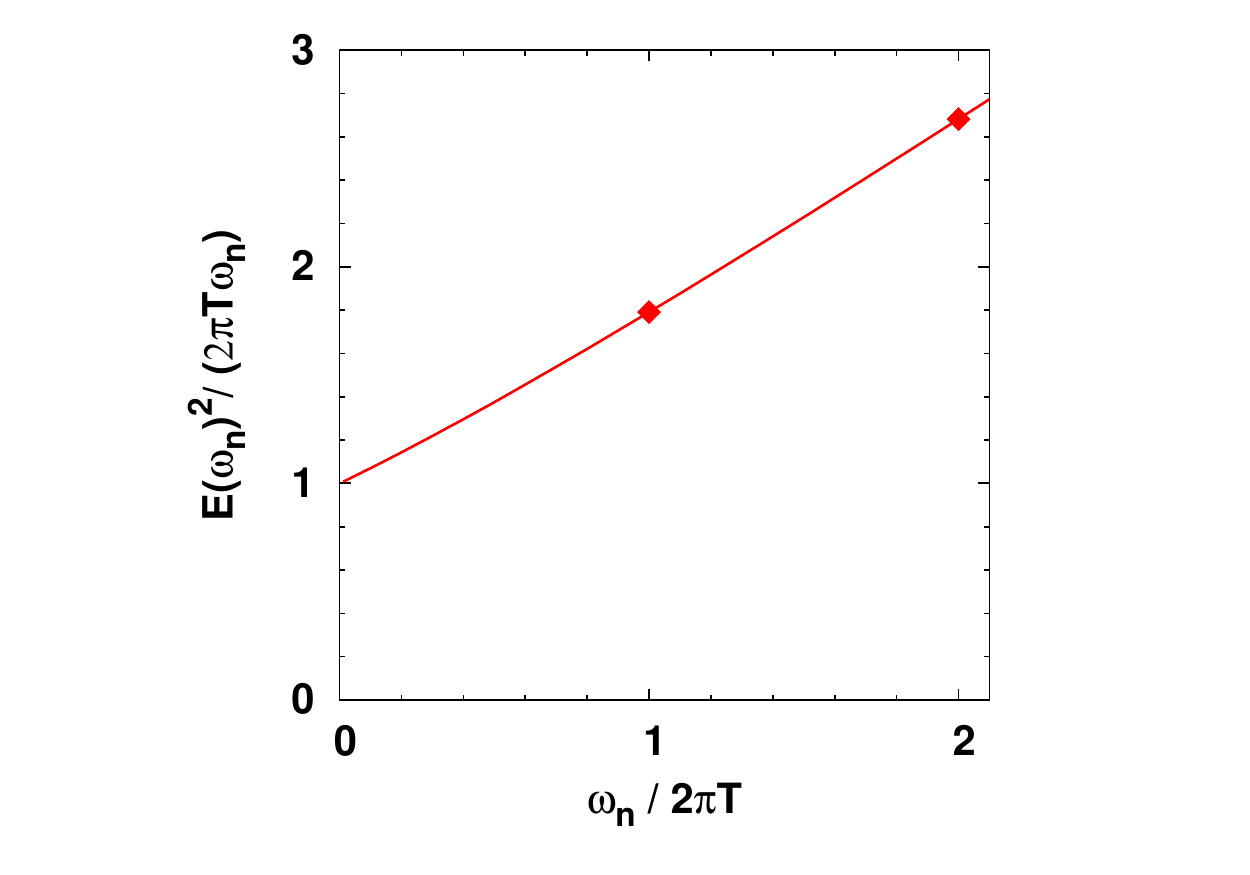}}
\vspace{-0.3cm}
\caption{Determination of the diffusion constant $D=1/(2\pi T)$ in AdS/CFT via the non-static screening masses
 related to $G^{(\omega_n)}_{00}$.}
\vspace{-0.8cm}
\label{fig:adscft}
\end{center}
\end{figure}

One example to illustrate the analytic continuation of the screening
masses in $\omega_n$ is provided by AdS/CFT. The ordinary differential
equation (ODE) that is solved to determine screening masses coupling
to a conserved current is essentially the same as the ODE solved to
obtain the spectral function describing the real-time excitations
coupling to the current; the difference resides in the boundary
conditions~\cite{Son:2002sd}. In the shear channel of the large-$N_c$, 
strongly coupled ${\cal N}=4$ SYM theory  for instance, Eq.\ (4.26) of
\cite{Kovtun:2005ev} for $\omega=0$ is equivalent to Eq.\ (2.20) of
\cite{Brower:1999nj}.
In the vector channel, starting from Eq.\ (4.5b) of \cite{Kovtun:2005ev},
we have to solve
\be\la{eq:Ez}
{\cal E}_z''(u) + \frac{\hat\omega^2 f'(u)}{(\hat\omega^2-\hat q^2 f(u)) f(u)} \;{\cal E}_z'(u) 
 + \frac{\hat\omega^2 -\hat q^2 f(u)}{u f^2(u)} \;{\cal E}_{z}(u) = 0
\ee
(${\cal E}_z$ can be interpreted as a component of the electric field to 
which the conserved vector current is coupled,
and $\hat q = \frac{q}{2\pi T}$,  $\hat\omega = \frac{\omega}{2\pi T}$).
In order to compute screening masses for Matsubara frequency $\omega_n$, we set 
$\hat\omega \doteq i \hat\omega_n$,  $\hat q \doteq  i \hat E(\hat\omega_n)$
in \eq(\ref{eq:Ez}) and choose the following boundary conditions:
(a) at $u=0$ (the boundary),  the solution ${\cal E}_z(u)$ must be normalizable, hence go to 0;
(b)  choose ${\cal E}_z(u)$ regular at the horizon $u=1$, i.e.\ $u^{+\hat\omega_n/2}$ 
rather than $u^{-\hat\omega_n/2}$ for $\omega_n>0$.
Using the shooting method starting at $u=1$, we compute the screening masses
for general positive $\omega_n>0$. As $\omega_n\to0$, we observe the behavior (\ref{eq:Esqrt}) 
with the known result $D=1/(2\pi T)$ \cite{Policastro:2002se} as expected, see \fig\ref{fig:adscft}.

\section{Conclusions}

`Integrating out' the non-static gauge modes perturbatively appears to
be a decent approximation even at $T=250$ MeV in the calculation of
screening masses.  Using a non-perturbative potential $V^+$ improves
the predictions for the non-static screening masses and
amplitudes. This study adds confidence to the applicability of EFT
methods for the study of phenomenologically interesting rate
observables at temperatures relevant to heavy ion collision
experiments.  An open question is whether the relation of non-static
screening masses and real-time rates can be extended to a
non-perturbative level. In the AdS/CFT context, a direct connection
exists between the calculation of screening masses and the calculation
of transport coefficients; whether this connection can be usefully
carried over to lattice calculations remains to be seen.

\bibliographystyle{elsarticle-num}
\bibliography{/Users/harvey/BIBLIO/viscobib.bib}

\end{document}